\documentclass[preprint2, 10pt]{aastex}
\usepackage{times, graphicx, natbib}
\usepackage[margin=1in]{geometry}

\slugcomment{}

\shorttitle{High Order Sunspot Oscillations}
\shortauthors{Jess et al.}

\begin{document}

\title{An Inside Look at Sunspot Oscillations with Higher Azimuthal Wavenumbers}

\author{David B. Jess$^{1,2}$, Tom Van Doorsselaere$^{3}$, Gary Verth$^{4}$, Viktor Fedun$^{5}$, S. Krishna Prasad$^{1}$, Robert Erd{\'{e}}lyi$^{4}$, Peter H. Keys$^{1}$, Samuel D. T. Grant$^{1}$, Han Uitenbroek$^{6}$ \& Damian J. Christian$^{2}$}
\affil{$^{1}$Astrophysics Research Centre, 
Queen's University Belfast, 
Belfast, BT7 1NN, 
UK}
\affil{$^{2}$Department of Physics and Astronomy,
California State University Northridge,
Northridge, CA 91330, 
USA}
\affil{$^{3}$Centre for mathematical Plasma Astrophysics, 
Department of Mathematics, KU Leuven,
Celestijnenlaan 200B bus 2400,
B-3001 Heverlee, 
Belgium}
\affil{$^{4}$Solar Physics and Space Plasma Research Centre (SP$^{2}$RC),
The University of Sheffield,
Hicks Building, Hounsfield Road,
Sheffield, S3 7RH, 
UK}
\affil{$^{5}$Space Systems Laboratory,
Department of Automatic Control and Systems Engineering,
University of Sheffield,
Sheffield, S1 3JD, 
UK}
\affil{$^{6}$National Solar Observatory$^\dagger$,
University of Colorado Boulder,
3665 Discovery Drive,
Boulder, CO 80303, 
USA}
\email{d.jess@qub.ac.uk}

\begin{abstract}
Solar chromospheric observations of sunspot umbrae offer an exceptional view of magneto-hydrodynamic wave phenomena. In recent years, a wealth of wave signatures related to propagating magneto-acoustic modes have been presented, which demonstrate complex spatial and temporal structuring of the wave components. Theoretical modelling has demonstrated how these ubiquitous waves are consistent with an $m=0$ slow magneto-acoustic mode, which are excited by trapped sub-photospheric acoustic ($p$-mode) waves. However, the spectrum of umbral waves is broad, suggesting that the observed signatures represent the superposition of numerous frequencies and/or modes. We apply Fourier filtering, in both spatial and temporal domains, to extract chromospheric umbral wave characteristics consistent with an $m=1$ slow magneto-acoustic mode. This identification has not been described before. Angular frequencies of $0.037 \pm 0.007~\mathrm{rad/s}$ ($2.1 \pm 0.4~\mathrm{deg/s}$, corresponding to a period $\approx$170~s) for the $m=1$ mode are uncovered for spatial wavenumbers in the range of $0.45<k<0.90$~arcsec$^{-1}$ ($5000-9000$~km). Theoretical dispersion relations are solved, with corresponding eigenfunctions computed, which allows the density perturbations to be investigated and compared with our observations. Such magnetohydrodynamic modelling confirms our interpretation that the identified wave signatures are the first direct observations of an  $m=1$ slow magneto-acoustic mode in the chromospheric umbra of a sunspot.
\end{abstract}


\section{Introduction} 
\label{sec:introduction}
\renewcommand{\thefootnote}{\fnsymbol{footnote}}
\footnotetext[2]{The National Solar Observatory is operated by the Association of Universities for Research in Astronomy under a cooperative agreement with the National Science Foundation.}
Since the early pioneering work by \citet{1969SoPh....7..351B}, \citet{1969SoPh....7..366W} and \citet{1970SoPh...13..323H}, to name but a few, oscillations and propagating waves tied to sunspot atmospheres have remained a challenging research area within solar physics. Observations have long indicated that wave power suppression exists in photospheric sunspot umbrae, with \citet{2007PASJ...59S.631N} providing a high-resolution view of this phenomenon with the Hinode/SOT instrument. Many theories have been put forward to explain such power suppression, including the absorption, scattering or channeling of field-guided magneto-acoustic waves following the mode conversion of $p$-mode oscillations \citep[e.g.,][]{1987ApJ...319L..27B, 1995ApJ...451..372C, 2003SoPh..214..201C, 2003MNRAS.346..381C,2016ApJ...817...45R}, the less efficient excitation of wave activity due to reduced turbulent convection \citep[e.g.,][]{1977ApJ...212..243G, 1988ApJ...326..462G}, and the reduction of attenuation lengths in the highly magnetic umbral regions of a sunspot \citep[e.g.,][]{1996ApJ...464..476J, 1997ApJ...476..392H}. What all of these theories have in common is the fact that the concentrated umbral magnetic fields modify the emerging wave signatures to produce magneto-acoustic wave activity \citep{2013AnGeo..31.1357Z}, which is observed to propagate anisotropically along the expanding magnetic field lines \citep[see the recent review articles by][]{2015SSRv..190..103J, 2016GMS...216..431V}. The interplay between various plasma measurements \citep[e.g., the magnetic field strength, the line-of-sight velocity, the intensity perturbations, etc.,][]{2009ApJ...702.1443F, 2014ApJ...791...61F, 2015A&A...578A..60M} has allowed researchers to verify that the majority of visible wave signatures in sunspot umbrae are synonymous with the $m=0$ slow magneto-acoustic mode. Indeed, such activity can readily be identified in chromospheric \citep[e.g.,][]{2007ApJ...671.1005B, 2007A&A...461L...1V, 2011A&A...525A..41K, 2013ApJ...779..168J, 2015arXiv150309106L, 2015A&A...579A..73M} and coronal \citep[e.g.,][]{2006RSPTA.364..461D, 2006A&A...448..763M, 2012ApJ...757..160J, 2016NatPh..12..179J, 2012A&A...546A..50K, 2015ApJ...812L..15K} sunspot-related studies involving both imaging and spectroscopic capabilities. 

Observations of sunspot umbral atmospheres often show increased activity as one moves away from the photospheric layer. \citet{2009ApJ...696.1683S} revealed evidence for dynamic filamentary structures in the chromosphere of a sunspot umbra when observed in the Ca~{\sc{ii}}~H absorption line. \citet{2013A&A...557A...5H} and \citet{2015A&A...574A.131H} found similar features, which were illuminated by the increased emission found in the vicinity of umbral flashes, suggesting there may be convective processes still at work within the cooler, magnetically dominated umbral atmosphere, allowing wave motion to more readily disturb the lower density chromospheric plasma \citep[see, also, the recent review by][]{2016GMS...216..467S}. This has important consequences, since it means that in the more-dynamic chromosphere, additional wave modes not readily identified (or suppressed) in the corresponding photosphere may present themselves more clearly. \citet{2001SoPh..203...71G} devised a static model atmosphere of a solar active region, and found that the plasma-$\beta$ (ratio of the plasma pressure to the magnetic pressure) was consistently less than unity across all atmospheric heights, indicating that the magnetic field will continue to play an important role in the propagation of waves through the chromosphere \citep[e.g.,][]{2014ApJ...792...41Y, 2016arXiv160105925L}, often creating radial structuring of the oscillation signals depending on the strength and orientation of the localized magnetic field, which are clearly visible in the Fourier power spectra maps presented by \citet{2012ApJ...746..119R} and \citet{2014A&A...569A..72S}.

\begin{figure*}[t!]
\figurenum{1}
\begin{center}
\includegraphics[width=0.9\textwidth]{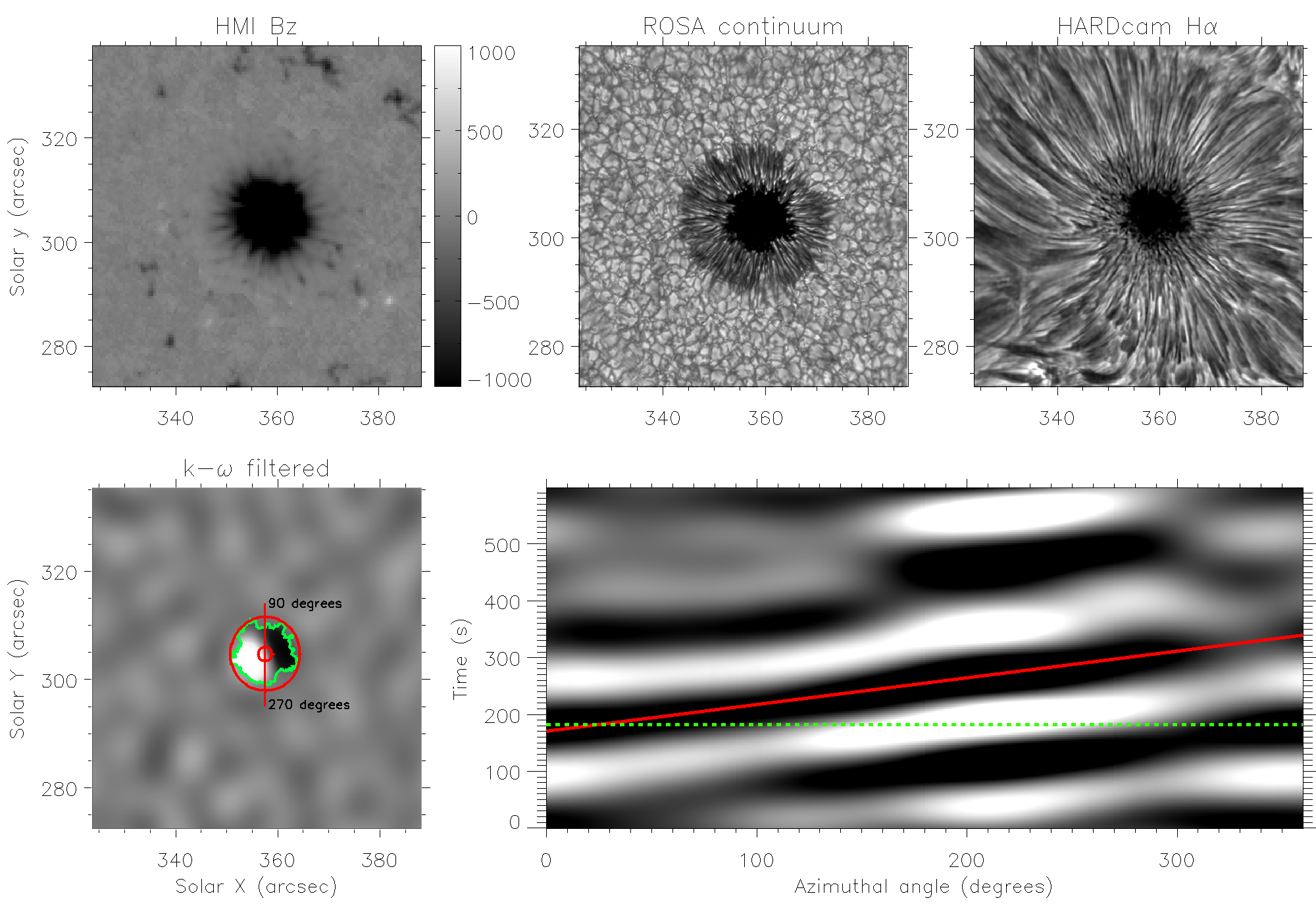}
\end{center}
\caption{Sample images of active region NOAA~11366 revealing the vertical component of the magnetic field (B$_{z}$; upper-left), the 4170{\,}{\AA} continuum (upper-middle) and the narrowband H$\alpha$ line core (upper-right). The color bar corresponding to the strength of the magnetic field is saturated at $\pm$1000{\,}G to better identify the sunspot structure. The lower-left panel displays a snapshot of H$\alpha$ intensities following both temporal and spatial frequency filtering. The green contour outlines the time-averaged umbra/penumbra boundary, while the red annulus depicts the extent of the region used for examining azimuthal wave motion within the umbra, where the center of the annulus is placed at the umbral barycenter. The lower-right panel is a time-azimuth diagram following the polar transformation of the signals contained within the red annulus in the lower-left panel, which allows the circular nature of the wave rotation to be investigated in a similar way to traditional time-distance diagrams. The horizontal dashed green line highlights the azimuthal intensity signal corresponding to the filtered image shown in the lower-left panel, which is also plotted in Figure~{\ref{fig_zoom_in}}, while the solid red line represents the fitted angular frequency of the rotating wave amplitudes.
\label{fig1}}
\end{figure*}

Modeling efforts focused on the excitation, propagation and/or dissipation of compressive waves in simplified solar atmospheres have been developed over a number of decades, with earlier examples including the work of \citet{1975SoPh...41..313C}, \citet{1977SoPh...54..269S} and \citet{1977A&A....54...61U}, to name but a few. More recent models have been constructed by, e.g., \citet{2006ApJ...653..739K}, \citet{2008SoPh..251..589K},  \citet{2011ApJ...727...17F}, \citet{2012ApJ...755...18V}, \citet{2013MNRAS.435.2589C}, \citet{2015A&A...577A..70S,2016A&A...590L...3S} and \citet{2017MNRAS.466..413C}. The excitation of longitudinal waves have been shown to be a consequence of the convective massaging of flux tubes \citep[magnetic pumping,][]{2011ApJ...730L..24K,2016ApJ...827....7K},  while on the other hand, \citet{2015ApJ...812L..15K} have observationally shown that their generation is rather connected to $p$-mode oscillations. The propagation of Alfv\'en waves in magnetic pores, and its potential for seismology, was described by \citet{2011ApJ...740L..46F}, \citet{2015MNRAS.449.1679M} and \citet{2015ApJ...799....6M}, with their observational signatures computed by \citet{2014PASJ...66S...9S}. The effect of neutrals on their dissipation and the resulting heating was further studied by \citet{2016ApJ...817...94A} and \citet{2016ApJ...819L..11S}.

Importantly, in recent years we have developed better imaging detectors that are more sensitive to incident photons. The benefits of this are twofold: (1) higher sensitivity equates to shorter exposure times, which helps to `freeze' atmospheric seeing when acquiring observations from ground-based facilities to help prevent spatial degradation, and (2) shorter exposure times allow for higher cadence image sequences, which raises the intrinsic Nyquist limit and allows us to probe high-frequency oscillations and propagating waves \citep[see Chapter~2 in the review article by][]{2015SSRv..190..103J}. Furthermore, better and more-robust adaptive optics systems are allowing longer duration studies of solar phenomena to be captured, providing a much improved frequency resolution for pinpointing and segregating particular oscillations of interest. Therefore, we are in an era where we can finally probe and examine the signatures and characteristics resulting from the superposition of multiple wave modes and harmonics within a single dataset. Here, in this article, we employ modern processing techniques to extract, interpret and model, for the first time, higher-order wave modes found within a sunspot umbral atmosphere.

\section{Observations \& Processing} 
\label{sec:observations}
The dataset used here has been thoroughly documented in previous studies \citep[e.g.,][]{2013ApJ...779..168J, 2016NatPh..12..179J, 2015ApJ...812L..15K}. However, for completeness, we will provide a brief overview. The image sequence duration was 75~minutes, and was obtained during excellent seeing conditions between 16:10 -- 17:25~UT on 2011 December 10 with the Dunn Solar Telescope (DST) at Sacramento Peak, New Mexico. The Rapid Oscillations in the Solar Atmosphere \citep[ROSA;][]{2010SoPh..261..363J} and Hydrogen-Alpha Rapid Dynamics camera \citep[HARDcam;][]{2012ApJ...757..160J} imaging systems were utilised to capture the near circularly-symmetric sunspot present within active region NOAA 11366, which was positioned at heliocentric co-ordinates (356{\arcsec}, 305{\arcsec}), or N17.9W22.5 in the conventional heliographic co-ordinate system. Here, we employ the blue continuum (52{\,}{\AA} bandpass filter centered at 4170{\,}{\AA}) and H$\alpha$ (0.25{\,}{\AA} filter centered on the line core at 6562.8{\,}{\AA}) filtergrams, with platescales of $0{\,}.{\!\!}{\arcsec}069$ and $0{\,}.{\!\!}{\arcsec}138$ per pixel, respectively, to provide a field-of-view size equal to $71{\arcsec}\times71{\arcsec}$. High-order adaptive optics \citep{2004SPIE.5490...34R} and speckle reconstruction algorithms \citep{2008A&A...488..375W} were implemented to improve the final data products, with final cadences of the continuum and H$\alpha$ channels equal to 2.11{\,}s and 1.78{\,}s, respectively. The Helioseismic and Magnetic Imager \citep[HMI;][]{2012SoPh..275..229S} present on the Solar Dynamics Observatory \citep[SDO;][]{2012SoPh..275....3P} provided simultaneous vector magnetograms of the active region with a cadence of 720{\,}s and a two-pixel spatial resolution of $1{\,}.{\!\!}{\arcsec}0$. A contextual HMI continuum image was also employed to co-align the images obtained from the DST with the full-disk HMI observations. Once aligned, a time-averaged 4170{\,}{\AA} continuum image was used to determine the umbral center-of-gravity, or intensity `barycenter', which forms the central co-ordinates of the umbral annulus used Section~{\ref{sec:discussion}}. Sample images of the data employed here are displayed in Figure~{\ref{fig1}}.

\begin{figure*}[t!]
\figurenum{2}
\begin{center}
\includegraphics[width=0.8\textwidth]{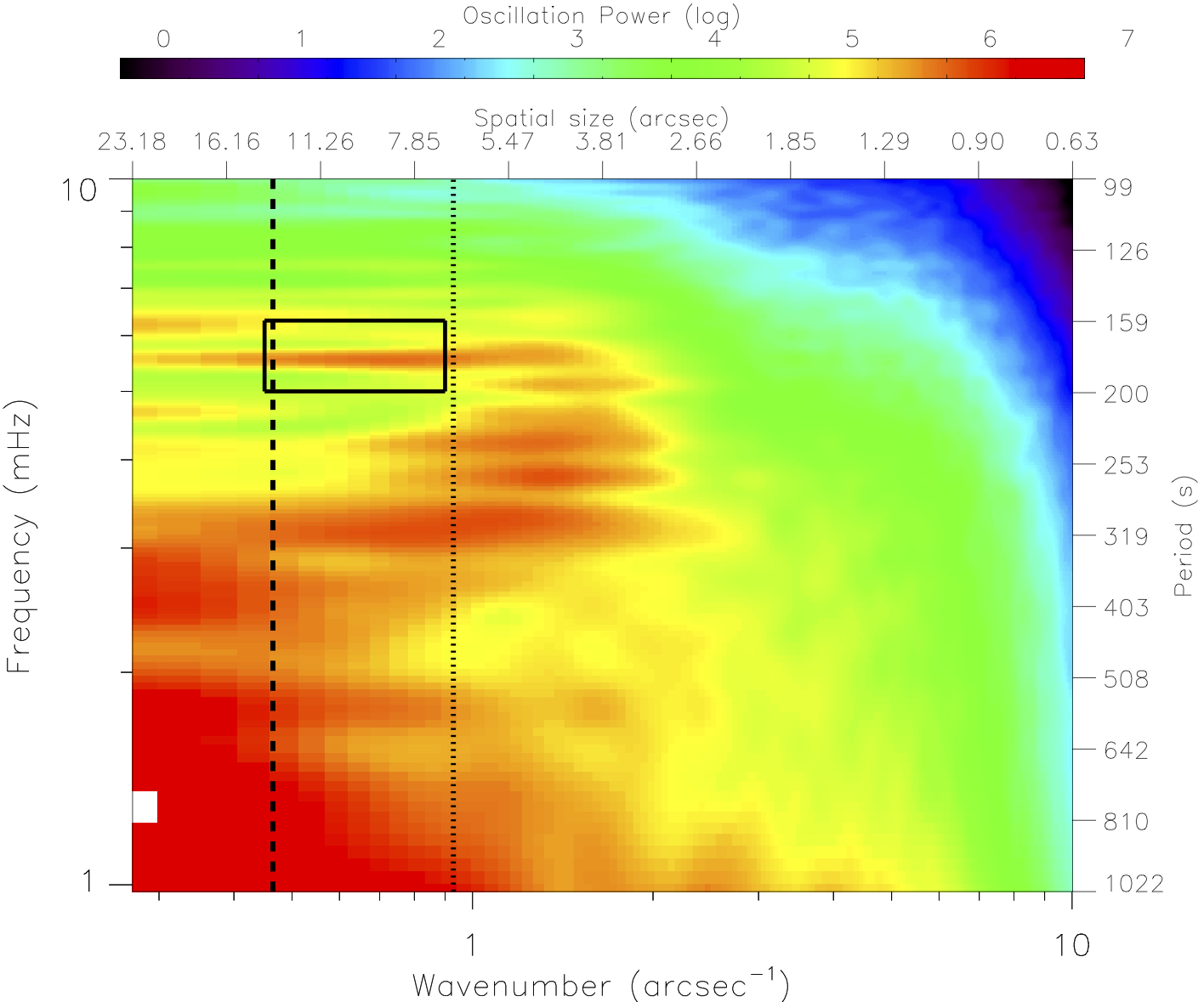}
\end{center}
\caption{A $k$-$\omega$ diagram, cropped to display spatial wavenumbers in the range of $0.27 < k < 10.02$~arcsec$^{-1}$ (23.18{\arcsec} -- 0.63{\arcsec}) and temporal frequencies in the range of $0.98 < \omega < 10.00$~mHz (99 -- 1022~s). The colors represent oscillatory power, shown on a log-scale, where red represents 7 orders-of-magnitude higher power than the background. The vertical dashed line corresponds to the spatial size of the umbral diameter ($\approx$13.5{\arcsec}), while the vertical dotted line represents the spatial size corresponding to the radius of the umbra ($\approx$6.75{\arcsec}). The solid black box highlights the FWHM of the chosen $k$-$\omega$ filter, which is seen to encapsulate a band of excess power at $\approx$170{\,}s over the entire spatial extent of the sunspot umbra.
\label{fig_komega}}
\end{figure*}

\section{Analysis \& Discussion}
\label{sec:discussion}
An image sequence obtained at the core of the H$\alpha$ line profile is employed to examine wave-related activity in the solar chromosphere for three distinct reasons. First, even with the reduced opacities present in the sunspot umbra, there is no evidence in our H$\alpha$ observations of umbral flash behaviour, which dominates datasets obtained at the core of the Ca~{\sc{ii}}~H/K or Ca~{\sc{ii}}~8542{\,}{\AA} absorption profiles. \citet{2017ApJ...sub..G} have shown statistically that umbral flashes first appear at an optical depth of $\log\tau \sim -3$, and preferentially manifest at an optical depth $\log\tau \sim -4.6$, corresponding to approximate geometrical heights of $\sim$250~km and $\sim$750~km, respectively \citep[sunspot model `M';][]{1986ApJ...306..284M}. Therefore, with this in mind, our H$\alpha$ umbral observations are likely to be formed at heights above 750~km, thus avoiding contamination from umbral flash events and making the visible intensity fluctuations purely related to the embedded (non-shocked) wave activity. Second, as the modelling efforts of \citet{2012ApJ...749..136L, 2013ApJ...772...90L} have revealed, the opacity of the H$\alpha$ line in the upper chromosphere is only weakly sensitive to the localised temperature, thus further reducing its sensitivity to high-forming umbral flash behaviour. Third, the time cadence of the H$\alpha$ observations is the highest (1.78~s), thus providing the best possible temporal frequency coverage, while still maintaining a diffraction-limited spatial resolution.

\begin{figure*}[t!]
\figurenum{3}
\begin{center}
\includegraphics[width=0.9\textwidth]{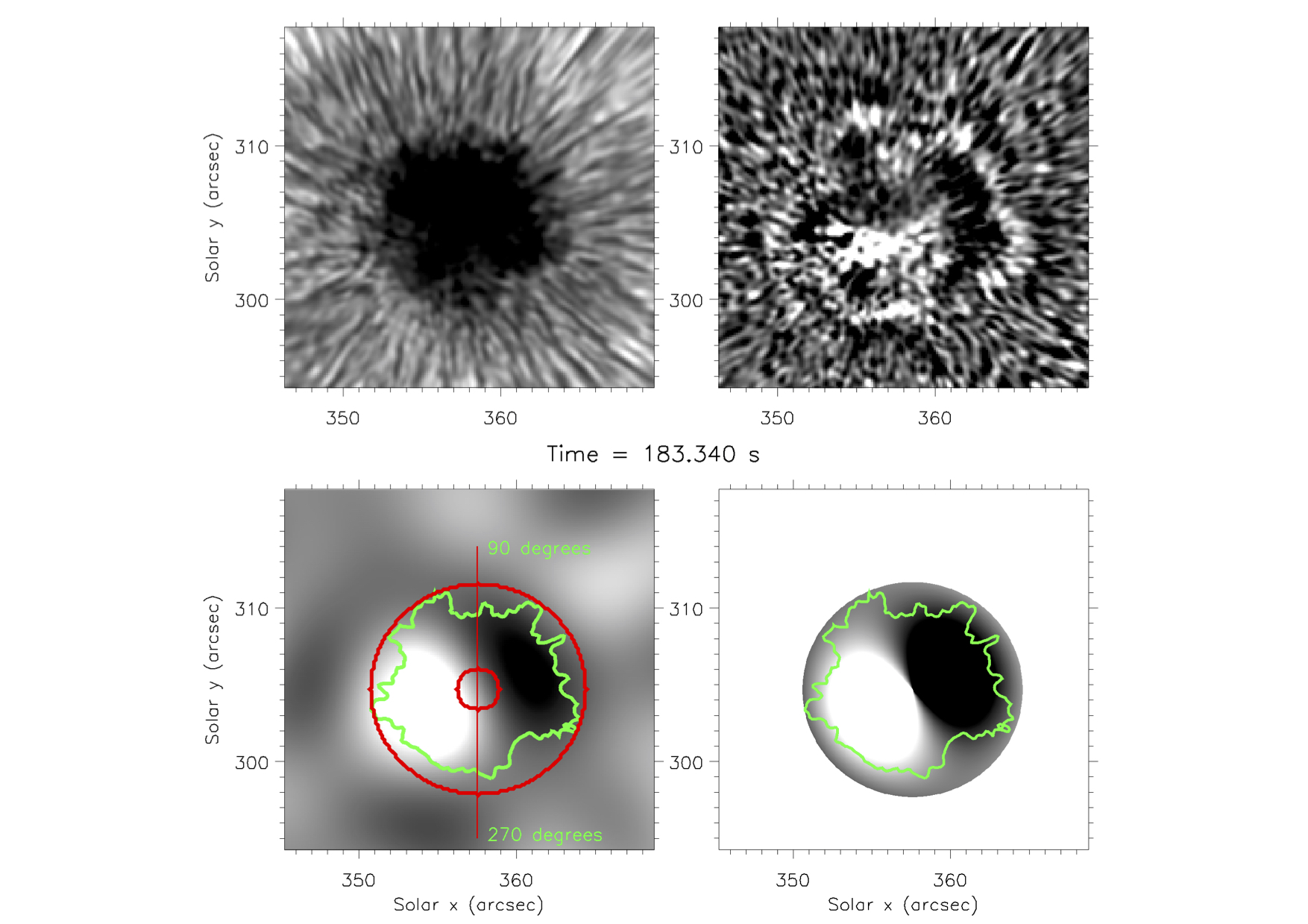}
\end{center}
\caption{A zoom-in of the unfiltered H$\alpha$ sunspot umbra (upper-left). The upper-right panel displays a simultaneous snapshot of the H$\alpha$ sunspot umbra having first been temporally filtered with a 160 -- 200{\,}s bandpass filter, while the lower-left panel reveals the simultaneous intensity fluctuations following the application of an additional 0.45 -- 0.90{\,}arcsec$^{-1}$ wavenumber filter, which reveals clear out-of-phase amplitude fluctuations at opposite edges of the sunspot umbra. The lower-right panel is the modelled density perturbation caused by an $m=1$ slow kink mode oscillation, which has been scaled to match the spatial size of the observed umbra for clarity. An animation of this Figure is available in the online edition.
\label{fig_zoom_in}}
\end{figure*}

The work of \citet{2013ApJ...779..168J} employed temporal filtering of the H$\alpha$ time series to provide a thorough understanding of dominant periodicities as a function of radial distance from the center of the umbra (or umbral `barycenter'). For the purposes of that work, no filtering was performed in the spatial domain. However, examining time-lapse movies of the temporally filtered H$\alpha$ images reveals a plethora of dynamic wave activity across a variety of spatial scales, particularly within the umbra where a dominant periodicity of $\sim$180{\,}s was uncovered, which is consistent with the work of \citet{2011A&A...525A..41K, 2013A&A...554A.146K, 2015SoPh..290..363K}. The enhanced oscillations, which are clearly observed in the temporally (150 -- 180{\,}s) filtered H$\alpha$ observations, are similar in magnitude to the outputs of chromospheric umbral resonance models put forward by \citet{1981SvAL....7...25Z} and \citet{1985SoPh...95...37S}, whereby the upwardly propagating slow magneto-acoustic waves, which do not violate the acoustic cut-off period \citep[e.g.,][to name but a few]{1977A&A....55..239B, 1991A&A...250..235F, 2008SoPh..251..501Z, 2014A&A...561A..19Y, 2015A&A...580A.107S}, are reflected continuously between the steep temperature gradients present close to the photospheric temperature minimum and at the transition region boundary. Of course, temperature and density gradients within the umbra provide a non-ideal resonator, which in turn gives wave amplitude inhomogeneities across the magnetic waveguide, similar to what is seen in the upper-right panel of Figure~{\ref{fig_zoom_in}}. These effects have been investigated previously by \citet{1988A&A...204..263L}, and more recently by \citet{2000PhDT........90N}, \citet{2000SoPh..192..403N}, \citet{2003ApJ...591..416C}, \citet{2011ApJ...728...84B} and the review article by \citet{2015LRSP...12....6K}, with observational evidence for such a scenario found by \citet{2015A&A...579A..73M}. However, importantly, the temporally filtered time series indicates that a component of the observed umbral oscillations are occurring on much-larger spatial scales than previously uncovered. 

To isolate and examine the presence of large-scale umbral oscillations, a complete $k$-$\omega$ filtering process was applied to the H$\alpha$ data, where $k$ is the spatial wavenumber (equal to $\frac{2\pi}{\lambda}$, where $\lambda$ is the spatial wavelength) and $\omega$ is the temporal frequency. Following the work of \citet{2013ApJ...779..168J}, a relatively broad temporal bandpass filter corresponding to 160 -- 200{\,}s (or 5.0 $<\omega<$ 6.3{\,}mHz) was employed to extract the dominant umbral oscillations. To examine the larger spatial fluctuations, a filter covering 7 -- 14{\arcsec} (or $0.45<k<0.90$~arcsec$^{-1}$) was chosen, as highlighed by the solid black box in Figure~{\ref{fig_komega}}. This wavenumber range was chosen since the diameter of the sunspot umbra is $\approx$98~pixels (see the outer edge of the annulus shown in Figure~{\ref{fig_zoom_in}}), corresponding to $\approx$13.5{\arcsec}, which means a spatial filter spanning 7 -- 14{\arcsec} will allow coherent oscillations of a similar size to the umbra to be investigated. Both the temporal and spatial filtering bandpasses are multiplied (in Fourier space) by a Gaussian envelope to reduce edge effects once transformed back into the space/time domain, hence the frequency ranges stipulated above are representative of the full-width at half-maximum of the corresponding $k$-$\omega$ filter. It is clear from Figure~{\ref{fig_komega}} that within the chosen $k$-$\omega$ filter there is a very strong oscillatory power signal, which is approximately 7 orders-of-magnitude above the background. The overall $k$-$\omega$ diagram depicts many of the quiet-Sun and internetwork features documented by \citet{2001A&A...379.1052K}, \citet{2011A&A...532A.111K} and \citet{2012ApJ...746..183J}, whereby higher temporal frequencies tend to be synonymous with larger spatial wavenumbers, producing the diagonal arm of enhanced oscillatory power seen in Figure~{\ref{fig_komega}}. However, within the boundaries of the applied $k$-$\omega$ filter, there is considerably elevated oscillatory power that spans a multitude of spatial scales (particularly within the range of $0.45<k<0.90$~arcsec$^{-1}$), yet remains relatively discrete in terms of the temporal frequency. This implies that the wave motion is best categorised by a narrow frequency range, yet demonstrates coherency across a broad spectrum of spatial scales, ranging from those close to the diameter of the sunspot umbra ($\approx$13.5{\arcsec}), through to those of similar size to the umbral radius, as indicated by the vertical dashed and dotted lines in Figure~{\ref{fig_komega}}, respectively. 

Oscillatory power, albeit reduced, is still clearly evident at smaller spatial wavenumbers than those associated with the umbral diameter. This implies that the discrete frequencies found within the umbra are still prevalent on much larger spatial scales, including outside the umbral waveguide. From the pioneering work of \citet{1970ApJ...162..993U} and \citet{1975A&A....44..371D}, which has subsequently been thoroughly developed by the use of modern, more sensitive instrumentation and techniques \citep[e.g.,][to name but a few]{1997SoPh..170...43K, 1997SoPh..175..287R, 1999ApJ...515..832H, RevModPhys.74.1073, 2004ApJ...608..562H, 2006ApJ...638..576G}, significant $p$-mode power at similar temporal frequencies (i.e., $\approx$3~minutes) has been found to coherently extend out to spatial wavelengths on the order of 100{\,}Mm ($\sim$140{\arcsec}), corresponding to wavenumbers $k \sim 0.05$~arcsec$^{-1}$. This is bigger than our current field-of-view, and indicates that large-scale coherent wave power readily exists in the solar photosphere at the temporal frequencies examined here. Of course, the H$\alpha$ observations presented in the current study are not only chromospheric in their composition (forming $\sim$1500{\,}km above the photosphere), but the presence of a highly magnetic sunspot embedded within the atmosphere naturally adds complexity to the picture \citep[see, e.g., the recent review by][]{2016GMS...216..489C}. Through multi-wavelength investigations, \citet{2010ApJ...721L..86R, 2013SoPh..287..107R} have demonstrated how sunspot structures can modify the observable characteristics of underlying 3~minute $p$-mode oscillations. Hence, a combination of chromospheric resonances and modified upwardly propagating $p$-mode oscillations may be the cause of the elevated wave power found at spatial scales exceeding that of the umbral diameter. Indeed, it seems likely that the observed heightened oscillatory power within the sunspot umbra may also be linked to the ubiquitous underlying $p$-mode oscillations. As per the work of \citet{2010ApJ...721L..86R, 2013SoPh..287..107R}, a multi-wavelength study (including photospheric observations) is necessary to examine the two-dimensional phase relationships with atmospheric height in order to conclusively verify whether the global $p$-modes are responsible for the observed wave power at large spatial scales. At spatial scales smaller than the umbral radius (i.e., $k>0.90$~arcsec$^{-1}$), the oscillatory power begins to decrease rapidly, while also shifting to slightly higher temporal frequencies in agreement with previously observed ($m=0$) magneto-acoustic wave phenomena \citep[e.g.,][]{2001A&A...379.1052K, 2011A&A...532A.111K, 2012ApJ...746..183J}. However, importantly, the lack of a positively correlated relationship between increasing $k$ and $\omega$ values within the chosen $k$-$\omega$ filter is not consistent with previous observations of traditional $p$-mode generated $m=0$ magneto-acoustic waves \citep[e.g.,][]{1988ApJ...324.1158D}, hinting at the presence of a more elusive wave mode.

Once the H$\alpha$ image sequence had been passed through the $k$-$\omega$ filter, it became very obvious from the resulting time series that large-scale spatially-coherent oscillations were manifesting within the chromospheric sunspot umbra, as indicated in the $k$-$\omega$ diagram displayed in Figure~{\ref{fig_komega}} and revealed in Figure~{\ref{fig_zoom_in}} (and the corresponding movie accessible in the online edition). The movie linked to Figure~{\ref{fig_zoom_in}} documents a 10-minute comparison between the unfiltered (raw), temporally filtered, and spatially {\it{and}} temporally filtered time series, in addition to the numerically modelled oscillations (see below), with simultaneous snapshots visible in the panels of Figure~{\ref{fig_zoom_in}}. Due to the azimuthal rotation about the umbral barycenter (central pivot of the annulus displayed in the lower-left panel of Figure~{\ref{fig_zoom_in}}, and visible in the associated movie), the intensities are averaged in the radial direction across the width of the annulus (40~pixels, or 5.5{\arcsec}). This is in agreement with the $k$-$\omega$ diagram presented in Figure~{\ref{fig_komega}}, whereby the frequency of oscillation contained within the chosen $k$-$\omega$ filter remains independent of the spatial scale (e.g., radius from the umbral barycenter), thus implying a rotation with constant angular frequency about the center of the sunspot umbra. Following the radial averaging of the umbral intensities, a polar transformation is performed to convert the azimuthal angle into a linearized array. Stacking these on top of one another produces the time-azimuth diagram shown in the lower-right panel of Figure~{\ref{fig1}}. Here, in a similar way to traditional time-distance diagrams, the gradients present in the time-azimuth panel relate to the rotational velocities, or more precisely, the angular frequencies (i.e., ${\mathrm{rad/s}}$ or ${\mathrm{deg/s}}$) of the wave mode. These are measured by following the techniques defined by \citet{2012NatCo...3E1315M} and \citet{2016NatPh..12..179J}, whereby a Gaussian profile is first fitted across the widths of the diagonal peaks (bright ridges) and troughs (dark ridges), before fitting a line-of-best-fit to the resulting Gaussian peaks and minimizing the sum of the squares of the residuals (i.e., least squares fitting; see the solid red line in the lower-right panel of Figure~{\ref{fig1}}). This provides angular frequencies of $0.037 \pm 0.007~\mathrm{rad/s}$ ($2.1 \pm 0.4~\mathrm{deg/s}$), corresponding to periodicities of $\approx$170~s, which (as expected) is within the range of the applied $k$-$\omega$ filter (160 -- 200{\,}s), yet more precisely quantifies the embedded temporal frequencies.

To model this wave, we consider the sunspot as a cylindrical structure in the polar coordinate system $(r,\phi,z)$, with the $z$-axis aligned with the umbral magnetic field, using the associated wave numbers $m$ and $k_z$ (following standard notation). We consider a plasma that changes its conditions (density, temperature, magnetic field) from the internal values (denoted with subscript `i') to the external values (denoted with subscript `e') with a step function. We take the centre of the sunspot as being a low-$\beta$ plasma, with a sound speed of $V_\mathrm{si}=6{\,}\mathrm{km/s}$ and Alfv\'en speed of $V_\mathrm{Ai}=12{\,}\mathrm{km/s}$. These choices are consistent with the Maltby `M' model used by \citet{2013ApJ...779..168J} for the same sunspot structure. Exterior to the sunspot, we consider an unmagnetised ($V_\mathrm{Ae}=0{\,}\mathrm{km/s}$) fluid with a sound speed $V_\mathrm{se}=9{\,}\mathrm{km/s}$. From the total pressure balance, we compute that the exterior is 3.4 times more dense than the interior of the sunspot. We now solve numerically the dispersion relation \citep[as derived by][]{zaitsev1975, 1983SoPh...88..179E} for slow waves in a cylindrical configuration. Given the observed behaviour, we take an azimuthal dependence of $m=1$. Moreover, we take $k_zR=30$ \citep[inspired by][]{2013A&A...555A..75M, 2016ApJ...817...44F}, where $R$ is the radius of the waveguide. For these parameters, we obtain a phase speed of $\omega/k_z=6.0{\,}\mathrm{km/s}$. Next, we computed the eigenfunctions of these waves in our assumed cylindrical configuration. To that end, we have used Eq.~14--23 of \citet{2016ApJS..223...23Y}, which relate the physical variables ($\rho$, $T$, $v_z$) to Bessel eigenfunctions. The main difference with the calculation in \citet{2016ApJS..223...23Y} is that all wave perturbations (in particular, the density and temperature) were put proportional to $\cos{(\omega t-\phi)}$. These perturbations were then added to the background density (assuming that the radial displacement of the oscillation is small, as would be expected for a slow-mode wave) to produce the image shown in the lower-right panel of Figure~{\ref{fig_zoom_in}}. Here, the radius of the cylindrical waveguide has been scaled, for clarity, to match the spatial size of the observed umbra, thus allowing a direct comparison to be made between the observed and simulated wave amplitudes displayed in the lower-left and lower-right panels, respectively, of Figure~{\ref{fig_zoom_in}}. In addition, the movie linked to Figure~{\ref{fig_zoom_in}} in the online edition displays the time evolution of the modelled $m=1$ slow magneto-acoustic mode, which is repeated continuously throughout the duration of the movie. Comparing the modelled wave signatures to those observed in our filtered observations reveals a remarkable level of consistency, further strengthening our interpretation that we have identified, for the first time, evidence for an $m=1$ slow magneto-acoustic wave propagating in the chromospheric umbra of a sunspot. 

While deriving the dispersion relation for the standard cylindrical case, we placed all perturbed variables proportional to $e^{i(k_zz +m\phi  -\omega t)}$. Considering the real part of the perturbations yields a displacement proportional to $\cos{(k_zz+\phi-\omega t)}$, in which $m=1$ for the kink asymmetry. When plotting these eigenfunctions as a function of time, they would be represented by anti-clockwise cork-screwing regions of high density along $\phi$ and $z$ that propagate upwards. Traditionally, for example in the case of coronal loop oscillations, we see propagating kink waves that oscillate in a plane. To model such instances, the solutions $\cos{(k_zz+\phi-\omega t)}$ and $\cos{(k_zz-\phi-\omega t)}$ are added together, which represent the $m=1$ and $m=-1$ modes, respectively. Simplification of the resulting motion would provide the displacement relation $\cos{(k_zz-\omega t)}\cos{(\phi)}$, where $\cos{(\phi)}$ is a steady state component no longer dependent on $t$ or $z$. Ultimately, adding the $m=1$ and $m=-1$ modes together produces a wave that only propagates in the $z$-direction (i.e., becomes a standing wave in the $\phi$ direction). In this case, however, we only consider the $m=1$ eigenfunction (i.e., not $m=-1$) as we wish to maintain the propagation behaviour in the $\phi$ direction (i.e., the apparent azimuthal motion). As our H$\alpha$ observations correspond to the upper chromospheric layer, it is not important for our present study whether the resulting wave is standing or propagating in the $z$-direction. Here, this distinction results from the superposition of independent waves with $\pm k_z$, and is something that will be investigated using simultaneous, multi-wavelength observations in a follow-up publication.

\begin{figure}[t!]
\figurenum{4}
\begin{center}
\includegraphics[width=\columnwidth]{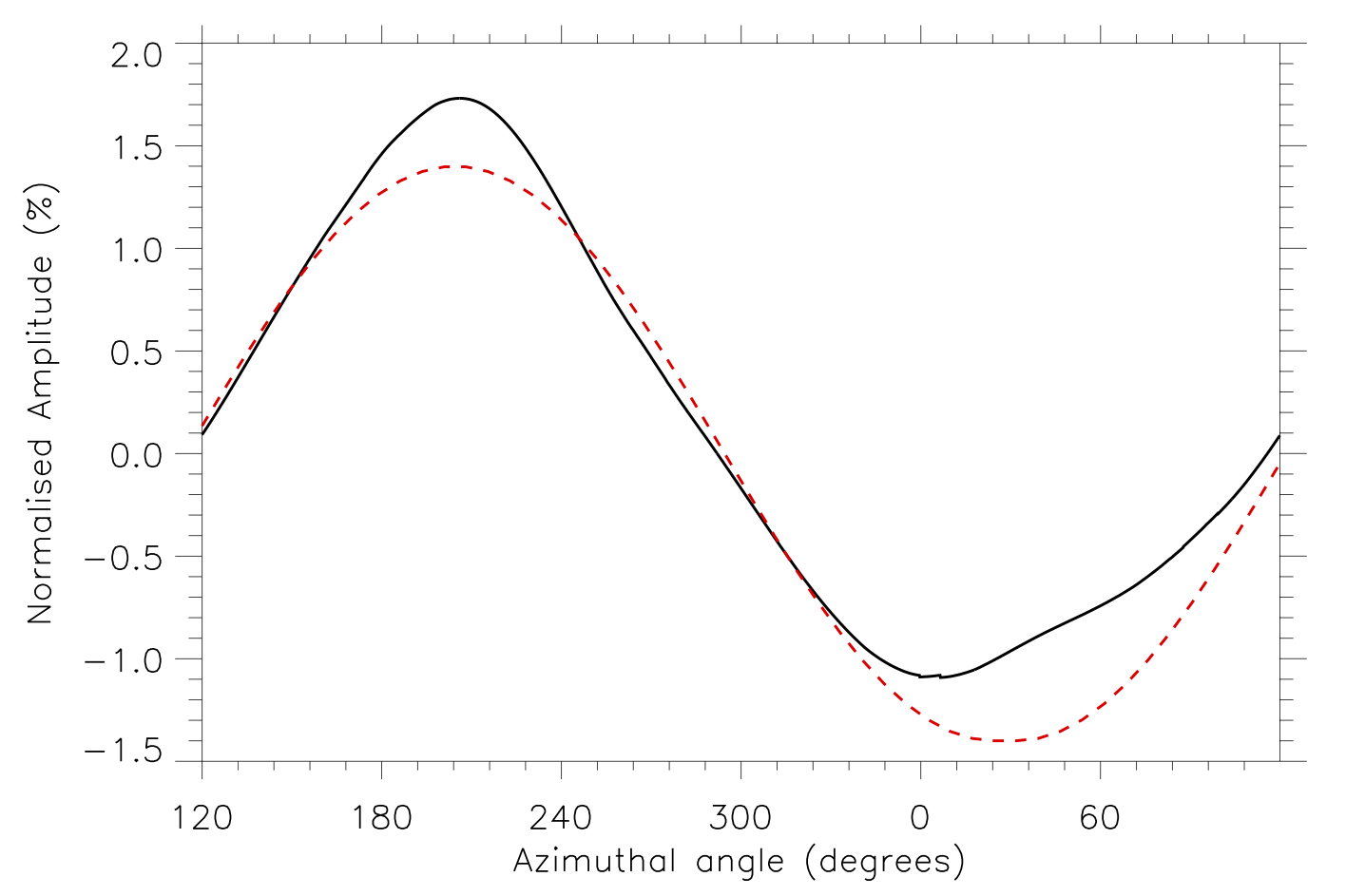}
\end{center}
\caption{The radially-averaged intensity fluctuations contained within the red annulus (lower-left panel of Figure~{\ref{fig_zoom_in}}; see also the dashed green line in the lower-right panel of Figure~{\ref{fig1}}) as a function of azimuthal angle around the umbral barycenter. The solid black line represents the amplitudes extracted from the observational H$\alpha$ data, while the dashed red line corresponds to the predicted amplitudes output from our cylindrical model for an $m=1$ slow magneto-acoustic kink mode.
\label{fig_amplitude_plot}}
\end{figure}

Alternatively, the observed angular frequency may be the consequence of the superposition of two perpendicularly polarised slow, kink waves, which are standing waves in their respective $\phi$ directions. Here, the initial conditions would require that two independent $m=\pm1$ slow kink waves are present, which are 90 degrees out-of-phase in $\phi$: (1) $\cos{(k_zz-\omega t)}\cos{(\phi)}$ and (2) $-\sin{(k_zz -\omega t)}\sin{(\phi)}$. The superposition of these two kink modes produces a density perturbation relation $\cos{(k_zz+\phi-\omega t )}$, which is identical to the fluctuations produced from a single, isolated $m=1$ slow kink mode. Therefore, while the driving mechanism for the observed wave behaviour may be different (e.g., a single, isolated $m=1$ slow kink mode or a pair of perpendicularly polarised $m=\pm1$ slow kink waves), the wave signatures produced (and observed) are identical.

It must be noted that while our observations clearly indicate apparent azimuthal motion related to the embedded density perturbations of the slow kink mode, this is distinctly different to the rotational twisting associated with torsional Alfv{\'{e}}n waves. In the case of Alfv{\'{e}}n waves, the physical bulk periodic rotation of magnetic field isocontours is a signature of such wave motion. Observationally, this may manifest as either the visible rotation of the magnetic feature (if well-resolved by the telescope), asymmetric Doppler velocities at opposite sides of the magnetic structure \citep[e.g.,][]{2014Sci...346D.315D, 2017NatSR...743147S}, or as periodic changes in the non-thermal line widths of the spectral lines used to observe the feature \citep[e.g.,][]{1998A&A...339..208B, 2009Sci...323.1582J}. For a more in-depth review, we refer the reader to the work of \citet{2009SSRv..149..355Z} and \citet{2013SSRv..175....1M}. However, in the case of the present analysis, no physical rotation of the sunspot (periodic or otherwise) is observed. Instead, we identify the bulk azimuthal rotation of Fourier power peaks inside the umbra, which are introduced by the density perturbations created from the presence of a single, isolated $m=1$ slow kink mode or a pair of perpendicularly polarised $m=\pm1$ slow kink waves. These signatures relate to the presence of the embedded wave mode (i.e., relative phase relationships across the spatial confines of the umbra), rather than a physical motion of the solar plasma. In addition, the signatures deduced in the present study are also distinctly different from those that would be associated with a fast kink mode. Here, the velocity components (i.e., the plasma flow field of the wave perturbations) are in the vertical direction, while a fast kink mode would be characterized by horizontal velocity perturbations. Furthermore, the intensity (i.e., density) fluctuations associated with fast kink modes would be significantly diminished as a result of the near incompressibility of these waves.

An interesting test to verify the robustness of our interpretation is to plot the instantaneous wave amplitudes as a function of azimuthal angle around the sunspot umbra, as defined by the annulus depicted in Figures~{\ref{fig1}} \& {\ref{fig_zoom_in}}. Following the polar transformation, the dashed green line in the lower-right panel of Figure~{\ref{fig1}} represents the instantaneous intensity fluctuations around the circumference of the annulus. These intensities are displayed in Figure~{\ref{fig_amplitude_plot}}, whereby a peak-to-peak amplitude is on the order of 2.8\% above the background, which is of the same order, albeit slightly smaller, as previous measurements of magnetically-confined slow-mode waves in the lower solar atmosphere \citep[e.g.,][]{2007A&A...473..943J, 2012A&A...544A..46B, 2015ApJ...806..132G}. The reduced peak-to-peak amplitude of the $m=1$ mode is likely a result of the relatively inefficient excitation mechanism for this mode, hence why the identification of such slow magneto-acoustic modes have proven impossible until the combination of modern high-resolution datasets and Fourier filtering techniques. Furthermore, as would be expected of an $m=1$ slow-mode wave, the intensity fluctuations, when plotted as a function of azimuthal angle, provide clear evidence of a single, well-resolved oscillation period. Figure~{\ref{fig_amplitude_plot}} displays both the observed (solid black line) and modelled (dashed red line) intensity fluctuations around the azimuth of the umbra. The similarities between the two curves highlight a continued consistency with our interpretation that we have identified an $m=1$ slow magneto-acoustic mode in the chromospheric umbra of our sunspot. Any slight misalignments between the modelled azimuthal fluctuations and those observed in our data may be the consequence of, for example, the non-perfect circular cross-section of the sunspot, shifts in the inclination angles of the umbral magnetic fields which affect the visible compressions of the localized plasma \citep[e.g.,][]{2014A&A...569A..72S, 2015ApJ...812L..15K}, or from changes in the opacity across the diameter of the sunspot which may modify the magnitude of the observed intensity fluctuations \citep[e.g.,][]{1965ApNr...10...17J, 2003ApJ...588..606K, 2014ApJ...795....9F}.

Of particular interest is the fact that the $m=1$ slow-mode wave is not omnipresent throughout the duration of the time series. The movie linked to Figure~{\ref{fig_zoom_in}} displays the visible manifestation, approximately three complete oscillation cycles, then the disappearance of the $m=1$ oscillation. This is in stark contrast to the ubiquitous $m=0$ slow-mode waves that thrive throughout all umbral time series. What is the reason behind this? Are the driving mechanisms completely different, therefore requiring special circumstances to induce the $m=1$ mode, which by itself may be a relatively inefficient wave driver? Or if driven by the underlying $p$-mode oscillations, could the broadness of this spectrum induce various $m=1$ modes at fractionally different angular frequencies, thus giving rise to beat phenomena that can modulate the signals produced by the (already weak) driver? Or, finally, could the not quite perfectly cylindrical shape of the sunspot umbra introduce slight differences between any $m=1$ and $m=-1$ eigenfunctions that might be present? Indeed, \citet{1999ApJ...518L.123N} found evidence for 3~minute magneto-acoustic oscillations surrounding the darkest central portion of an irregularly shaped sunspot umbra, although the Fourier power maps presented did not allow any temporal variability to be investigated. Perhaps, in such a regime, additional beating of these two modes (on top of what might be present from $m=1$ modes at fractionally different angular frequencies) might occur, thus introducing a quasi-periodic nature of the observed wave phenomenon. 

\section{Conclusions}
\label{sec:conclusions}
Here, we have presented high spatial and temporal resolution H$\alpha$ observations, captured by the HARDcam instrument at the Dunn Solar Telescope, of wave activity in the umbra of a sunspot. On the date of the observations, 2011 December 10, the sunspot corresponding to active region NOAA 11366 was very quiet and exhibited near-circular geometry. Within the immediate vicinity of the sunspot, a $k$-$\omega$ diagram revealed the traditional trend of lower-frequency oscillations being associated with larger spatial scales \citep[e.g., as detailed in][]{2013ApJ...779..168J}. However, of particular interest was a region of high oscillatory power, which corresponded to a constant frequency ($\approx$5.9{\,}mHz, $\approx$170{\,}s or $\approx$0.037{\,}$\mathrm{rad/s}$) over a wide range of spatial wavenumbers ($0.45<k<0.90$~arcsec$^{-1}$ or 7 -- 14{\arcsec}). Through the application of a $k$-$\omega$ filter, this oscillation was isolated and further studied. 

Through modelling the sunspot as a cylindrical structure in the polar coordinate system $(r,\phi,z)$, with the $z$-axis aligned with the umbral magnetic field, we solved the intrinsic dispersion relation for an $m=1$ slow mode wave and computed the corresponding eigenfunctions. We find that the modelled density perturbations remain consistent with our high-resolution observations, suggesting we have uncovered a large-scale isolated $m=1$ slow kink mode oscillation in the chromospheric umbra of a sunspot. However, through analysis of the mathematical eigenfunctions, our observations may also be consistent with a pair of perpendicularly polarised $m=\pm1$ kink waves. While the wave signatures produced will be identical, the underlying driving mechanism may be vastly different; something that will require further study utilizing a plethora of multiwavelength observations. Thus, for the first time, we have presented a detailed examination of slow kink mode oscillations in the chromospheric umbra of a sunspot, which display spatial coherency on distances of up to 14{\arcsec} (or $k\approx0.45$).

\acknowledgments
DBJ thanks the UK Science and Technology Facilities Council (STFC) for an Ernest Rutherford Fellowship, in addition to a dedicated standard grant that allowed this project to be undertaken. DBJ also wishes to thank Invest NI and Randox Laboratories Ltd. for the award of a Research \& Development Grant (059RDEN-1) that allowed the filtering techniques employed to be developed.
TVD was supported by an Odysseus grant of the FWO Vlaanderen, the IAP P7/08 CHARM (Belspo) and the GOA-2015-014 (KU~Leuven).
GV, VF, SKP and RE also wish to thank the UK STFC. 
RE further acknowledges support by the Chinese Academy of Sciences President's International Fellowship Initiative, Grant No. 2016VMA045 and the Royal Society (UK).
PHK is grateful to the Leverhulme Trust for the award of an Early Career Fellowship that allowed this work to be undertaken. 
SDTG wishes to thank the UK Department of Employment and Learning for a PhD studentship.
DJC is grateful to CSUN for start-up funding.
The contextual imaging and magnetic field measurements employed in this work are courtesy of NASA/SDO and the AIA, EVE, and HMI science teams.


\end{document}